\documentclass{appolb}
\usepackage{graphicx}
\usepackage{ogonek}
\usepackage{amsmath}
\usepackage{bm}

\begin{document}
\title{On introducing Charge-Symmetry-Breaking terms\\ to nuclear Energy Density Functionals
\thanks{Presented at the XXXVI Mazurian Lakes Conference on Physics, Piaski, Poland, September
1-7, 2019.}%
}
\author{
P. B\k{a}czyk$^a$,
M. Konieczka$^a$,
K.M.L. Martinez$^b$,
S. Anti\'{c}$^b$,
P.A.M~Guichon$^c$,
W. Satu\l{}a$^a$,
J.R. Stone$^{d,e}$,
A.W. Thomas$^b$
\address{
$^a$Institute of Theoretical Physics, Faculty of Physics, University of Warsaw, ul.~Pasteura 5, PL-02-093 Warsaw, Poland\\
$^b$CSSM and CoEPP, Department of Physics, University  of Adelaide, SA~5005~Australia\\
$^c$IRFU-CEA, Universit\'{e} Paris-Saclay, F91191 Gif sur Yvette, France\\
$^d$Department of Physics (Astro), University of Oxford, OX1 3RH United Kingdom\\
$^e$Department of Physics and Astronomy, University of Tennessee, TN 37996 USA\\
}
} 

\maketitle
\begin{abstract}
The Charge-Symmetry-Breaking (CSB) character of the nucleon-nucleon interaction is well established. This work presents two different ways of introducing such effects into a nuclear Energy Density Functional (EDF). CSB terms are either coming from the effective theory expansion or are derived from electromagnetic mixing of $\rho^0$ and $\omega$ mesons. These terms are then introduced to Skyrme and Quark-Meson-Coupling EDFs, respectively.
\end{abstract}

\section{Introduction}

An atomic nucleus is typically described as a many-body system of point-like nucleons. Its existence is determined by a delicate balance between the repulsive Coulomb interaction and the effective nuclear force. However, the origin of the latter interaction as well as basic properties of nucleons (like mass and charge) can be understood only by increasing the resolving power of the description and carefully studying quarks confined in nucleons. Because of their different quark structure, proton (\textit{uud}) and neutron (\textit{udd}) have different charges and masses and interact in a slightly different way~\cite{Mil06}. In particular, it is well established that the strengths of $pp$ and $nn$ interaction in the same spin channel are not equal~\cite{Mac01}. This fact is known as Charge Symmetry Breaking (CSB).

One of the main goals of nuclear physics is to formulate a theoretical framework able to track down the influence of confined quarks on the properties of atomic nuclei. Solving the quantum chromodynamics (QCD) problem of quarks and gluons using calculations on the lattice is currently limited to systems with only few nucleons and even there the description is model-dependent~\cite{Fuk95, Bea12, Yam12}. Lately, the chiral effective field theory ($\chi$EFT) has become a popular tool to derive the \textit{NN} interaction including its CSB character~\cite{Mac16}. In parallel, the \textit{NN} force can be modelled with a great precision by phenomenological potentials, e.g.\ AV18~\cite{Wir95}, which are fitted to the experimental data on low-energy scattering processes. Both types of interaction can be then incorporated into \textit{ab initio} nuclear many-body techniques to obtain detailed calculations of atomic nuclei. Unfortunately, the applicability of those methods is often limited to proximity of magic shell closures or by a certain mass number due to the numerical complexity.

A more universal approach is provided by models based on the mean-field approximation, e.g.\ Density Functional Theory (DFT), however, they cannot be used with the realistic \textit{NN} potentials due to the numerical problems resulting from the hard-core repulsion. A successful Energy Density Functional can be built starting with Skyrme interaction~\cite{Sky56} which results from effective theory expansion. Recently, an EDF coming from quark-meson-coupling (QMC) model has been proposed~\cite{Gui04, Gui06, Sto16}. In this way, a concept of nucleons being ensembles of quarks interacting with mesonic exchanges is transferred into the EDF which can be then used in global DFT calculations. Worth noting is that both Skyrme and QMC EDFs, as defined in their original form, have no terms which violate charge symmetry. As such they cannot reproduce the main signature of CSB in atomic nuclei, i.e.\ Mirror Displacement Energies defined as $\mathrm{MDE}=\mathrm{BE}\left(A,\ T,\ T_z=-T\right)-\mathrm{BE}\left(A,\ T,\ T_z=+T\right)$, where $\mathrm{BE}<0$ is binding energy of a nucleus, $A$ is the mass number, $T$ and $T_z = \frac12(N-Z)$ are the total isospin and its $z$ component. Although MDEs depend predominantly on the CSB effects introduced by the Coulomb force, a contribution of this kind coming from the effective nuclear force is needed to reproduce the experimental data~\cite{Nol69}.

In this work, two different approaches of introducing CSB terms to these EDFs are presented. First, the Skyrme interaction is extended in the spirit of effective theory. Second, the CSB term modelling electromagnetic mixing of $\rho^0$ and $\omega$ mesons is added to the QMC EDF. The comparison of both methods is also provided.

\section{Adding CSB terms to Skyrme EDF}

The contact force proposed by Skyrme~\cite{Sky56} is derived in the scope of effective theory. The short-range \textit{NN} interaction is expanded as a series in momentum transfer and only terms up to the next-to-leading order are kept and Fourier transformed into coordinate space. In this way the interaction on the shortest distances, irrelevant to low-energy nuclear structure, is approximated in a systematic way. Additionally, the density-dependent and spin-orbit terms are considered. Altogether this interaction is characterized by up to 11 adjustable parameters. Considering applications, Skyrme-based EDFs~\cite{Vau72} in various parametrizations prove their worth in reproducing bulk properties of atomic nuclei among the entire nuclear chart. Recently, in order to study MDEs, the CSB extensions to the isoscalar Skyrme interaction have been added in leading order (LO)~\cite{Bac18}:
\begin{equation}\label{eq:V_III}
\hat{V}_0^\mathrm{III} = t_0^\mathrm{III}   \left( \hat{\tau}^{}_z + \hat{\tau}_z' \right) \delta \left( r \right).
\end{equation}
and next-to-leading order (NLO)~\cite{Bac19} of the effective theory. The form of the isospin dependence is taken from the Henley-Miller classification~\cite{Hen79}. The free parameters of the model (one coupling constant in LO and three coupling constants in NLO) are fitted to all available experimental data on MDE in isospin doublets ($A=7$ to $75$) and isospin triplets ($A=6$ to $58$). Different parametrizations of the isoscalar part of the Skyrme EDF are used. It is shown that, already in LO, the class-III interaction is sufficient to reproduce experimental MDEs in both isospin doublets and triplets~\cite{Bac18} and that the addition of the NLO terms improves the accuracy of calculated MDEs of the lightest and the heaviest multiplets. What is more, it turns out that the newly-added contact terms model in fact CSB introduced by effective nuclear interaction~\cite{Bac19}. This conclusion is drawn from a comparison of the results obtained within DFT and the calculations based on the AV18 potential~\cite{Wir95} performed with \textit{ab initio} Green Function Monte Carlo (GFMC) method~\cite{Car15}.

\section{Adding CSB terms to QMC EDF} 

Another approach of constructing an EDF for nuclear calculations has been proposed recently in Ref.~\cite{Sto16}, where authors derived the functional from the quark-meson-coupling (QMC) model~\cite{Gui88, Gui96, Gui04, Gui06}. In this formulation nucleons are treated as confined, non-overlapping bags of three quarks. Then, the nuclear interaction is described with exchange of mesons between quarks from different bags. The adjustable parameters of the model are coupling constants of $\rho$, $\omega$ and $\sigma$ mesons and experimentally uncertain mass of the $\sigma$ resonance. Since its first appearance in 2006, the QMC EDF has undergone a handful of important modifications. First, the pion exchange was introduced to the model to properly account for the \textit{NN} interaction on longer ranges. Lately, the EDF was improved by adding the tensor term predicted by the model. 

Recently, the functional dubbed QMC$\pi$-0 has been implemented into the robust numerical code \texttt{HFODD}~\cite{Sch17}. The adjustable parameters of the model were fitted to the experimental data on magic and semi-magic nuclei. It was verified that such an approach provides results which are comparable with those obtained with modern parametrizations of Skyrme EDFs. The publication on details concerning the form of the functional, fitting procedure and obtained results is being prepared~\cite{Kon20}.

Similarly to the Skyrme EDF, the QMC EDF requires additional CSB terms to account for MDEs. In terms of meson exchanges the electromagnetic mixing of $\rho^0$ and $\omega$ mesons is often suggested to be the main source of CSB effects~\cite{Coo87, Mil90}.\footnote{The CSB effects in nuclear interaction can be also solely attributed to nucleon mass splitting and its influence on two-pion exchange. The consensus on this issue is not yet established~\cite{Con97, Mac01b, Mil06}.} Such a mesonic contribution to the EDF can be derived starting with \textit{NN} scattering matrix in the momentum space~\cite{Hen79}. For small momentum transfers, $q$, the dominant term reads:
\begin{equation}\label{eq:expansion}
\mathcal{M}_{NN}^{(\rho^0-\omega)}(q) = \frac{g_\rho g_\omega \langle \omega | H_{em} | \rho^0 \rangle} {m_\rho^2 m_\omega^2} \left(\hat{\tau}_z^{}+\hat{\tau_z}'\right) + \ldots,
\end{equation}
where $g_\rho=\sqrt{G_\rho}m_\rho$, $g_\omega=\sqrt{G_\omega}m_\omega$ and $\langle \omega | H_{em} | \rho^0 \rangle$ is the electromagnetic mixing element of $\rho^0$ and $\omega$ mesons. After Fourier transforming the interaction to the coordinate space one obtains:
\begin{equation}
V_{NN}^{(\rho^0-\omega)}(r) = \frac{\sqrt{G_\rho G_\omega} \langle \omega | H_{em} | \rho^0 \rangle} {m_\rho m_\omega} \left(\hat{\tau}_z^{}+\hat{\tau}_z'\right) \delta(r).
\end{equation}
Finally, such a potential can be compared with the class-III interaction in LO introduced to the isoscalar Skyrme force~\eqref{eq:V_III}. Matching terms in front of the $\delta$ function leads to:
\begin{equation}\label{eq:t3_mixing}
t_{0,\rho\omega}^\mathrm{III} = \frac{\sqrt{G_\rho G_\omega} \langle \omega | H_{em} | \rho^0 \rangle} {m_\rho m_\omega} = -6.7(6)\,\mathrm{MeV\,fm}^3. 
\end{equation}
The numerical values are as follows: $G_\rho = 4.90$\,fm$^2$, $G_\omega = 6.92$\,fm$^2$ (taken from the adjustment of the QMC$\pi$-0 EDF to experimental values of masses and charge radii of magic nuclei~\cite{Kon20}), $m_\rho = 3.90$\,fm$^{-1}$, $m_\omega = 3.95$\,fm$^{-1}$ (as in the \texttt{HFODD} code used for calculations), $\langle \omega | H_{em} | \rho^0 \rangle = -3500(300)\,\mathrm{MeV}^2$ (taken from Ref.~\cite{Gar98}). In this way, the coupling constant of the term modelling $\rho^0$-$\omega$ mixing was not adjusted to any CSB effects in finite nuclei. Interestingly, its value given in eq.~\eqref{eq:t3_mixing} is very close to values obtained for Skyrme EDFs with the CSB term in LO: $t_{0,\text{SV}_\text{T}}^\mathrm{III} = -7.3(3)\,\mathrm{MeV\,fm}^3$, $t_{0,\text{SkM*}}^\mathrm{III} = -5.4(2)\,\mathrm{MeV\,fm}^3$ and $t_{0,\text{SLy4}}^\mathrm{III} = -5.5(2)\,\mathrm{MeV\,fm}^3$~\cite{Bac18,Bac19,BacPhD}.

\begin{figure}[t!]
\centerline{%
\includegraphics[scale=1.0]{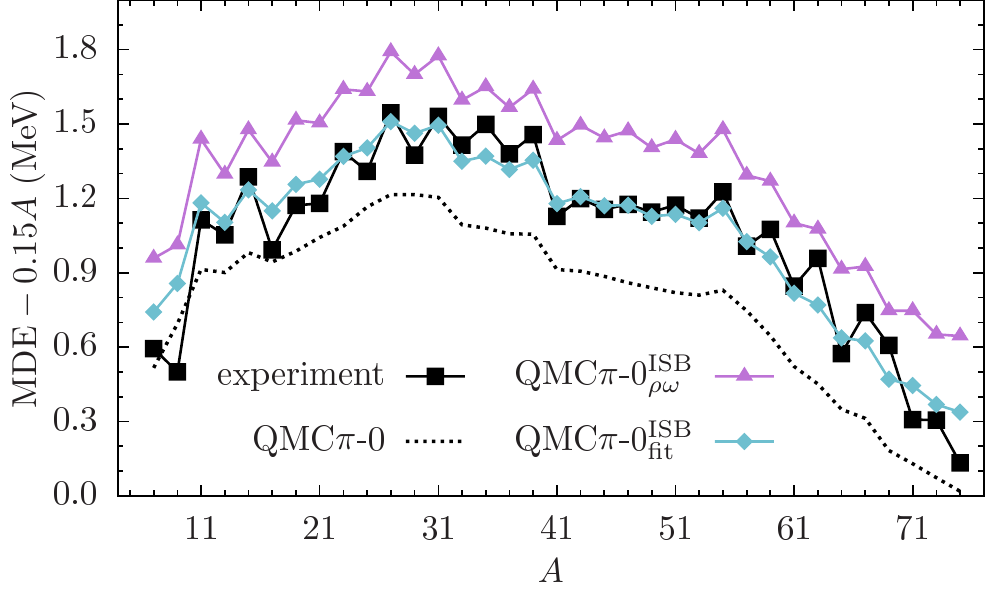}}
\caption{MDEs for $T=\frac12$ mirror nuclei calculated using QMC$\pi$-0$^\mathrm{CSB}$ with $t_0^\mathrm{III}$ parameter coming from the model of $\rho^0-\omega$ mixing (violet triangles) or from the fit (turquoise diamonds). The results are compared with the experimental values~\cite{Wan17} (black squares) and the calculations for QMC$\pi$-0 EDF (black dashed line).}
\label{Fig:QMC}
\end{figure}

The parameter $t_0^\mathrm{III}$ can be also adjusted to reproduce the experimental values of MDE as described in Refs.~\cite{Bac18,Bac19}. For the QMC$\pi$-0 EDF one obtains $t_{0,\mathrm{fit}}^\mathrm{III} = -3.5(3)\,\mathrm{MeV\,fm}^3$, which provides a very good agreement with the experimental data as shown in Fig.~\ref{Fig:QMC} presenting the calculated MDEs for isospin doublets for both values of $t_0^\mathrm{III}$ parameter. Systematically too large CSB contribution coming from the term modelling $\rho^0$-$\omega$ mixing may be a result of neglecting either higher order terms in the expansion~\cite{Mac01} or some other important mesonic CSB mechanisms. 

\section{Conclusions}

In this contribution, two ways of introducing CSB terms into different EDFs were discussed. The Skyrme EDF was successfully extended with an effective class-III interaction~\cite{Bac18,Bac19} while the QMC EDF was augmented with a term coming from $\rho^0-\omega$ mixing mechanism. The calculated coupling constant of the latter modification turned out to be twice as large as the coupling constant required to reproduce the experimental MDEs. Possible ways of reducing this discrepancy are implementing higher-order terms of expansion in eq.~\eqref{eq:expansion} or other mesonic effects contributing to CSB.

This work was supported in part by the Polish National Science Centre
under Contract Nos.~2014/15/N/ST2/03454, 2015/17/N/ST2/04025, 2017/24/T/ST2/00159, 2017/24/T/ST2/00160 and~2018/31/B/ST2/02220. We acknowledge CI\'S \'Swierk Computing Center, Poland, for the allocation of computational resources. J.R.S. and P.A.M.G. acknowledge with pleasure the support and hospitality of CSSM at the University of Adelaide during visits in the course of this project. This work was supported by the University of Adelaide and by the Australian Research Council through the Australian Research Council (AU) Centre of Excellence for Particle Physics at the Terascale (CE110001104) and Discovery Projects No. DP150103101 and No. DP180100497.


\begin{thebibliography}{99}

\bibitem{Mil06}{G.A. Miller \textit{et al.}, Annu. Rev. Nucl. Part. Sci. \textbf{56}, 253 (2006).}

\bibitem{Mac01}{R. Machleidt, Phys. Rev. C \textbf{63}, 024001 (2001).}

\bibitem{Fuk95}{M. Fukugita \textit{et al.}, Phys. Rev. D \textbf{52}, 3003 (1995).}

\bibitem{Bea12}{S.R. Beane \textit{et al.}, Phys. Rev. D \textbf{85}, 054511 (2012).}

\bibitem{Yam12}{T. Yamazaki \textit{et al.}, Phys. Rev. D \textbf{86}, 074514 (2012).} 

\bibitem{Mac16}{R. Machleidt and F. Sammarruca, Physica Scripta \textbf{91}, 083007 (2016).}

\bibitem{Wir95}{R.B. Wiringa \textit{et al.}, Phys.Rev. C \textbf{51}, 38 (1995).}

\bibitem{Sky56}{T.H.R. Skyrme, Phil. Mag. \textbf{1}, 1043 (1956).}

\bibitem{Gui04}{P.A.M. Guichon and A.W. Thomas, Phys. Rev. Lett. \textbf{93}, 132502 (2004).}

\bibitem{Gui06}{P.A.M. Guichon \textit{et al.}, Nucl. Phys. A \textbf{772}, 1 (2006).}

\bibitem{Sto16}{J.R. Stone \textit{et al.}, Phys. Rev. Lett. \textbf{116}, 092501 (2016).}

\bibitem{Nol69}{J.A. Nolen and J.P. Schiffer, Annu. Rev. Nucl. Sci. \textbf{19}, 471 (1969).}

\bibitem{Vau72}{D. Vautherin and D.M. Brink, Phys. Rev. C \textbf{5}, 626 (1972).}

\bibitem{Bac18}{P. B\k{a}czyk \textit{et al.}, Phys. Lett. B \textbf{778}, 178 (2018).}

\bibitem{Bac19}{P. B\k{a}czyk \textit{et al.}, J. Phys. G: Nucl. and Part. Phys. \textbf{46}, 03LT01 (2019).}

\bibitem{Hen79}{E. Henley and G.A. Miller, \textit{Mesons in nuclei}, (North-Holland, 1979).}

\bibitem{Car15}{J. Carlson \textit{et al.}, Rev. Mod. Phys. \textbf{87}, 1067 (2015).}

\bibitem{Gui88}{P.A.M. Guichon, Phys. Lett. B \textbf{200}, 235 (1988).}

\bibitem{Gui96}{P.A.M. Guichon \textit{et al.}, Nucl. Phys. A \textbf{601}, 349 (1996).}

\bibitem{Sch17}{N. Schunck \textit{et al.}, Comput. Phys. Commun. \textbf{216}, 145 (2017).}

\bibitem{Kon20}{M. Konieczka, P. B\k{a}czyk, K.M.L. Martinez, P.A.M. Guichon, J. Stone, A.W.~Thomas, in preparation.}

\bibitem{Coo87}{S.A. Coon and R.C. Barrett, Phys. Rev. C \textbf{36}, 2189 (1987).}

\bibitem{Mil90}{G.A. Miller \textit{et al.}, Phys. Rep. \textbf{194}, 1 (1990).}

\bibitem{Con97}{H.B. O'Connell \textit{et al.}, Prog. Part. Nucl. Phys. \textbf{39}, 201 (1997).}

\bibitem{Mac01b}{R. Machleidt and H. M\"{u}ther, Phys. Rev. C \textbf{63}, 034005 (2001).}


\bibitem{Gar98}{S. Gardner \textit{et al.}, Phys. Rev. Lett. \textbf{80}, 1834 (1998).}

\bibitem{BacPhD}{P. B\k{a}czyk, \textit{PhD thesis}, University of Warsaw (2019).}

\bibitem{Wan17}{M. Wang \textit{et al.}, Chin. Phys. C \textbf{41}, 030003 (2017).}

%
%
%
%
%
%
%
%
%
%
%
%
%
%
%
%
%
%
%
%
%
%
%
%
%
%
%
%

\end{thebibliography}
\end{document}